\documentstyle[epsf,rotate]{mn}

\title[Thermal stability and nova cycles in permanent superhump systems]
{Thermal stability and nova cycles in permanent superhump systems}

\author[A. Retter \& T. Naylor]
   {A.~Retter
   and T. Naylor\\
  Department of Physics, Keele University, Keele, Staffordshire, 
  ST5 5BG, UK; ar@astro.keele.ac.uk; timn@astro.keele.ac.uk\\}

\date{Accepted ???. Received ???; in original form ???}

\begin{document}

\maketitle

\begin{abstract}

Archival data on permanent superhump systems are compiled to test the 
thermal stability of their accretion discs. We find that their discs are 
almost certainly thermally stable as expected. This result confirms 
Osaki's suggestion (1996) that permanent superhump systems form a new 
subclass of cataclysmic variables (CVs), with relatively short orbital 
periods and high mass transfer rates. We note that if the high accretion 
rates estimated in permanent superhump systems represent their mean secular 
values, then their mass transfer rates cannot be explained by gravitational 
radiation, therefore, either magnetic braking should be extrapolated to 
systems below the period gap or they must have mass transfer cycles. 
Alternatively, a new mechanism that removes angular momentum from CVs 
below the gap should be invoked.

We suggest applying the nova cycle scenarios offered for systems above 
the period gap to the short orbital period CVs. Permanent superhumps 
have been observed in the two non-magnetic ex-novae with binary periods 
below the gap. Their post-nova magnitudes are brighter than their 
pre-outburst values. In one case (V1974~Cyg) it has been demonstrated that 
the pre-nova should have been a regular SU~UMa system. Thus it is the first 
nova whose accretion disc was observed to change its thermal stability.
If the superhumps in this system indicate persistent high mass transfer 
rates rather than a temporary change induced by irradiation from the hot 
post-nova white dwarf, it is the first direct evidence for mass transfer 
cycles in CVs. The proposed cycles are driven by the nova eruption.

\end{abstract}


\section{Introduction}

\subsection{Regular superhumps}

Superhumps were initially observed in SU~UMa systems (Vogt 1974; Warner 
1975). This subclass of dwarf novae, shows brighter longer outbursts (so 
called superoutbursts) in addition to the normal outbursts observed in 
regular dwarf novae (U~Gem systems). A quasi-periodic variation, 
systematically a few per cent larger than the orbital period, is usually 
detected during SU~UMa superoutbursts, and was nicknamed `$\bf superhump$' 
(see la Dous 1993; Warner 1995a for reviews of SU~UMa systems and CVs in 
general). The typical value of the peak-to-trough amplitude of the 
superhump is about 20-40 per cent. There is a trend of increasing 
positive period excess (of the superhump period over the orbital one) 
with the orbital period (Stolz \& Schoembs 1984; Patterson 1999).

Vogt (1982) first suggested that the accretion disc around the white 
dwarf in Z Cha (and other SU~UMa systems) develops an elliptical shape 
during superoutbursts. Osaki (1985) tested the motion of single particles 
in an eccentric disc, and found a relation between the superhump period 
excess as a fraction of the binary period and the  binary period. 
Whitehurst (1988) used hydrodynamic simulations to show that the tidal 
instability explains the formation of the eccentric disc. Osaki (1989) 
used two instabilities in the accretion disc to create a uniform theory 
for non-magnetic CVs. According to this model, `The Thermal -- Tidal 
Disc Instability Model', the superhump periodicity is the beat of the 
orbital period of the binary system with the period of the apsidal 
precession of the accretion disc. Further hydrodynamical simulations 
showed that the tidal instability can occur only if the disc radius 
exceeds a certain value, the 3:1 resonance radius (Whitehurst \& King 
1991). This requires that superhumps can appear only in binary systems 
with a small mass ratio q=($M_{2}/M_{1})$$\la$0.33, although 
observationally the limit is probably bigger (Retter \& Hellier 2000; 
Retter et al. 2000).

During the last decade it has been found that superhump behaviour is 
common among other classes of binary systems as well as in the SU~UMa 
stars. According to the theory, there are two requirements for the 
presence of superhumps: an extreme mass ratio and a large accretion disc 
radius. These two conditions are naturally met in superoutbursts of 
SU~UMa systems, but are satisfied in other systems as well. Thus 
superhumps have also been detected in the ultra-short orbital period 
AM~CVn systems (Patterson et al. 1993a; Patterson, Halpern \& Shambrook 
1993; Provencall et al. 1997; Patterson et al. 1997a; Harvey et al. 
1998; Solheim et al. 1998; Patterson 1999; Skillman et al. 1999); in 
SW~Sex stars (Patterson \& Skillman 1994; Patterson 1999) and during 
bright outbursts of a few low-mass X-ray transients (White 1989; Charles 
et al. 1991; Bailyn 1992; Zhang \& Chen 1992; Kato, Mineshige \& Hirata 
1995; O'Donoghue \& Charles 1996).


\subsection{Permanent superhumps}

In many systems, which do not show eruptions, superhumps are observed. 
An outburst is thus not a necessary condition for this phenomenon. 
Patterson \& Richman (1991) termed this behaviour as `$\bf permanent$ 
$\bf superhump$'. During the last few years permanent superhumps have 
been found in almost twenty CVs (see Patterson 1999 for an observational 
review on permanent superhump systems). Typical full amplitudes of 
permanent superhumps are about 5-15 per cent, but they are highly 
variable, and sometimes even disappear from the light curve (Patterson, 
personal communication). Their name is thus somewhat misleading.

In several systems, a quasi-stable periodicity, a few per cent shorter 
than the orbital period has been observed. These variations are called 
`$\bf negative$ $\bf superhumps$'. In a few systems they appear 
simultaneously with the positive superhumps (Patterson et al. 1997b; 
Arenas et al. 2000; Retter \& Hellier 2000; Retter et al. 2000), but in 
other cases they are the only kind of superhump observed. An alternation 
between the positive and negative superhumps has been observed in a few 
objects (e.g. Skillman et al. 1998). The negative superhump deficit over 
the orbital period seems to be correlated with the orbital period in a 
manner similar to the Stolz \& Schoembs (1984) relation, however with 
a shallower trend. It has been suggested that negative superhumps are 
formed by the precession of the accretion disc in the azimuthal axis 
(Patterson et al. 1993c; Patterson 1999), but there are some theoretical
difficulties with this idea (Murray \& Armitage 1998; Wood, Montgomery \&
Simpson 2000). 


Osaki (1996) further proposed that only the values of the orbital period 
and the mass transfer rate determine the basic differences among the four 
major subclasses of non-magnetic CVs (U~Gem systems, SU~UMa stars, 
permanent superhumpers and nova-likes). Fig.~1 is taken from his paper. 
The permanent superhump systems are thought to be in a state of higher 
accretion rate (implying large accretion disc radii) than in regular 
SU~UMa systems. Osaki's model describes pretty well most superhump 
observations, although his suggestion that the period gap separates 
between systems with tidally unstable accretion discs and tidally stable 
objects is violated by the presence of many permanent superhump systems 
above the gap (Patterson 1999; Retter \& Hellier 2000; Retter et al. 2000; 
see also Table~1 and Fig.~2).

\begin{figure}

\centerline{\epsfxsize=3.0in\epsfbox{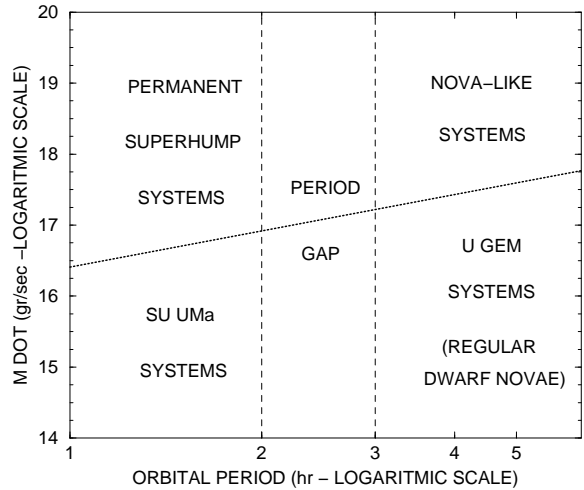}}

\caption{A $\bf schematic$ theoretical diagram based on Osaki (1996), 
showing the location on the (orbital period, mass transfer rate) plane 
of four major groups of non-magnetic CVs. The two dashed vertical lines 
represent the two ends of the period gap in the period distribution of 
CVs. Systems on the right hand side have accretion discs that are tidally 
stable. CVs on the left are tidally unstable. Objects above the dotted 
tilted line have thermally stable discs. Systems below that line are 
thermally unstable.}

\end{figure}

\subsection{Superhumps in classical nova systems}

The permanent superhump model has been invoked so far for three classical 
novae: V603~Aql (Patterson \& Richman 1991; Patterson et al. 1993c; 
Patterson et al. 1997b), CP~Pup (White \& Honeycutt 1992; White, Honeycutt 
\& Horne 1993; Thomas 1993; Patterson \& Warner 1998) and V1974~Cyg 
(Retter, Leibowitz \& Ofek 1997; Skillman et al. 1997). In all cases an 
alternative magnetic explanation has been proposed as well (Haefner \& 
Metz 1985; White et al. 1993; Balman, Orio \& Ogelman 1995; Semeniuk et 
al. 1995; Olech et al. 1996), however the arguments for the superhump 
explanation seem much stronger. The periods of the three novae fit well 
in the Stolz \& Schoembs (1984) diagram, and in two cases negative 
superhumps have been observed in addition to the presence of positive 
superhumps (Patterson et al. 1997b; Patterson 1999). Nova V4633 Sgr 1998 
is another permanent superhump candidate (Lipkin, personal communication;
see also Lipkin \& Leibowitz 2000).

Retter \& Leibowitz (1998) introduced a simple way of testing the thermal
stability state of accretion discs in CVs. Employing this method on the 
three permanent superhump novae they found that these systems are indeed 
thermally stable, while the progenitor of V1974~Cygni was located below 
the critical line for stability. This result led Retter \& Leibowitz to 
suggest that if the decline from outburst in V1974~Cyg towards the pre-nova 
magnitude continues, the post-nova should evolve into a regular SU~UMa 
system with superhumps appearing only during superoutbursts. Alternatively, 
its disc might stay optically thick, keeping above the thermal stability 
limit, and continue to show superhumps permanently in its light curve. 
Retter \& Leibowitz thus proposed that non-magnetic classical novae can 
be progenitors of permanent superhump systems. Retter, Naylor \& Leibowitz 
(1999) and Retter \& Naylor (2000) developed this proposal. In this work 
we elaborate and extend these ideas.

\subsection{Existing nova cycle models for CVs}

The different subclasses of CVs share a similar configuration, namely 
a binary system with a primary white dwarf and a Roche-lobe filling 
secondary red dwarf. Spectra of old novae usually show strong continua 
and emission lines, very similar to nova-like systems (Bode \& Evans 
1989; Warner 1995a). A few observations further support a possible 
connection between classical novae and dwarf novae. Two old novae 
experience regular dwarf nova outbursts a few decades after their 
eruption -- the peculiar nova, GK~Per 1901 (Sabbadin \& Bianchini 1983), 
and V446~Her 1960 (Honeycutt, Robertson \& Turner 1995; Honeycutt et al. 
1998). Dwarf nova outbursts a few decades before the nova eruption of 
V446~Her have probably been observed as well. Livio (1989) and Warner 
(1995a) listed a few other cases, however, the evidence for such a 
transition in these systems is rather poor. 

Robinson (1975) compared the magnitudes of 18 old novae with the values 
of their progenitors. He found no significant difference between these 
numbers, and concluded that all novae return to their pre-outburst 
luminosities. The observational properties of old novae are very similar 
to those of nova-likes, which have thermally stable accretion discs 
(Warner 1995b). Therefore, it seems that the outburst does not alter 
the thermal stability at all, at least for time scales of the order of 
a few decades. An exception in Robinson's sample is V446~Her, mentioned 
above. It is, however, the only clear case of a regular nova\footnote{We 
exclude GK~Per as it is a very atypical nova -- its orbital period is 
very long ($\sim$2 d), and its secondary star is believed to be a 
sub-giant (e.g. Dougherty et al. 1996) unlike red dwarf companions in 
classical novae.} that had dwarf nova outbursts a few decades after (and 
probably even before) the nova eruption. Recently, it was found that V446~Her 
has an orbital period near 5 h (Thorstensen \& Taylor 2000) -- typical for 
classical novae.

Vogt (1990) investigated a sample of 97 old novae. His results seem to 
confirm the idea that there is no systematic difference in the brightness 
of pre-novae and post-novae. He also showed that the brightness of old 
novae tends to fade slowly in the decades following their outbursts. This 
finding was confirmed by another study (Duerbeck 1992). A possible 
interpretation of the decrease in the nova light is a future transition 
to a different phase (dwarf nova).
 
Therefore it was suggested that nova outbursts link different CV 
subclasses. The nova cycle of the subgroups of CVs is, however, still 
debatable, and several scenarios have been offered for the connections 
among these subclasses.

The `hibernation scenario' (Shara et al. 1986; Prialnik \& Shara 1986; 
Shara 1989) suggests that dwarf novae $\rightarrow$ nova-likes 
$\rightarrow$ novae $\rightarrow$ nova-likes $\rightarrow$ dwarf novae 
$\rightarrow$ `hibernation' $\rightarrow$ dwarf novae etc. However, it 
was later proposed that the `hibernation' phase ($\dot{M}$$\approx$0) 
might not exist at all (Livio 1989), thus dwarf novae $\rightarrow$ 
nova-likes $\rightarrow$ novae $\rightarrow$ nova-likes $\rightarrow$ 
dwarf novae... The typical time scales for the transitions were estimated 
as a few centuries -- millenia. 

An alternative view to the `hibernation scenario' and to the `modified / 
mild / modern hibernation scenario' was presented by Mukai and Naylor 
(1995). They suggested that nova-likes and dwarf novae constitute different 
classes of pre-nova systems. Therefore, there are two possibilities:
1. nova-likes $\rightarrow$ novae $\rightarrow$ nova-likes...
2. dwarf novae $\rightarrow$ novae $\rightarrow$ dwarf novae...
Nova-likes should have more frequent nova outbursts than dwarf novae 
because their mass transfer rates are larger than those of dwarf novae. 
The critical mass for the thermonuclear runaway is thus achieved much 
faster. Transitions between the two phases are allowed on the long term
scale.

It seems that the observations of old novae have not been able to judge 
between the various models (Naylor et al. 1992). Furthermore, these
scenarios were suggested when there were essentially no known non-magnetic
novae below the period gap, and before the discovery of the permanent 
superhump class. In this work we extend the models to the short orbital 
period CVs and test them by the observations accumulated so far.

\section{The location of the permanent superhump systems in the 
($\bf P_{orb}$, $\bf \dot{M}$) plane}

To test Osaki's suggestion (1996), that the accretion discs in permanent 
superhump systems are thermally stable (unlike the discs of SU~UMa systems 
that are unstable in quiescence, and become quasi-stable only during 
superoutbursts), we locate these systems in the ($P_{orb},\dot{M}$) plane. 
We basically follow the method developed by Retter \& Leibowitz (1998). 
It essentially uses $m_{V}$ to estimate $\dot{M}$, and assumes that the 
accretion disc is the dominant light source in the V band. It is also
assumed that the disc is not kept at a high temperature due to irradiation 
by the white dwarf, which is relatively hot in post-novae. A modification 
that we add to these calculations is the effect of the inclination angle, 
i, on the visual magnitude, expressed by Warner (1987): 

\begin{equation}
\bigtriangleup M_{i}=-2.5 log[(1+1.5cos(i))cos(i)]
\end{equation}

where $\bigtriangleup M_{i}$ is the magnitude change as a function of i.
The resulting equations equivalent to equations (7) (for the critical 
instability value) and (8) (for the calculation of accretion rates) 
of Retter \& Leibowitz (1998) are:

\begin{equation}
(m_{V})_{crit}=2.16-4.25logP_{orb}-3.33logM_{1}+5logd+A_{V}-
\bigtriangleup M_{i}
\end{equation}

\begin{equation}
\dot{M_{17}}=(10^{\frac{m_{V}+\bigtriangleup M_{i}-A_{V}-0.69}{-2.5}})
\frac{d^{2}}{M_{1}^{4/3}}
\end{equation}

where our symbols are identical with those of Retter \& Leibowitz.

In Table~1, we present values of the relevant parameters of permanent 
superhump systems compiled from various sources. The list of objects is 
primarily based on Patterson (1999). Only `conventional' permanent superhump 
systems (i.e. only positive superhumpers) were chosen. Systems showing only 
negative superhumps in their light curves were rejected from our sample. 
Note that evidence for the presence of positive superhumps in TV~Col has 
been presented by Retter \& Hellier (2000) and Retter et al. (2000). The 
permanent superhump AM~CVn systems were also excluded. The mass transfer 
rates were calculated by equation (3) using the values in the table as 
input parameters.

\begin{table*}
\begin{minipage}{300 mm}
\caption{Parameters of the permanent positive superhump systems}
\begin{tabular}{@{}lccccccc}
Object  & $P_{orb}$ & $M_{wd}$          &     d             &     $A_{V}$       &   $m_{V}$         &      i          & $\dot{M}$ \\
Name/s  &    (h)    &($M_{\odot}$)      &   (kpc)           &                   &                   &                 & (10$^{17}$gr/sec)\\
PX~And= &3.51$^{1}$ & 0.70$^{1}$        &    $>0.18^{1}$    &0.15-0.17$^{1}$    &14.9-15.0$^{1}$    & 74$^{1}$        & $>0.04$ \\
PG 0027+260&        &                   &                   &                   &                   &                 &         \\

V603~Aql= &3.32$^2$ & 0.66-1.40$^{3,4}$ & 0.33-0.38$^{3}$   &0.22-0.5$^{5,6}$   & 11.2-12.0$^{7}$   &13-20$^{2,3,4}$  & 5.8-59.7 \\
Nova Aql 1918&      &                   &                   &                   &                   &                 &         \\

TT~Ari  &3.30$^{8}$ & 0.79$^{9}$        &0.285-0.385$^{10}$ &0.15-0.17$^{11,12}$&9.5-11$^{13,14,15}$& 15-35$^{9,11}$  & 17.5-168 \\

V592~Cas&2.76$^{16}$& 0.8-1.44$^{17}$   & 0.063$^{18}$      &0.45-0.8$^{16,19}$ & 12.6-12.8$^{16}$  & 18-39$^{17}$    & 0.09-0.44 \\

TV~Col=& 5.49$^{20}$ & 0.6-0.9$^{20}$   & 0.5-0.6$^{21}$    & 0.18-0.20$^{21}$  & 13.6-14.1$^{21}$  & 30-72$^{20,21}$ & 0.65-11.1 \\
2A 0526-328&        &                   &                   &                   &                   &                 &         \\

V1974~Cyg=&1.95$^{22}$&0.89-1.07$^{23,24,25}$&1.66-1.88$^{26}$& 0.96-1.02$^{26}$& 15.9-16.1$^{27}$  & (36-54)$^{26}$  & 4.6-15.6 \\
Nova Cyg 1992&      &                   &                   &                   &                   &                 &         \\

V795~Her=&2.60$^{28}$& 0.7$^{29}$       &                   & 0.12-0.13$^{30}$  & 12.2-12.4$^{31}$  & 56$^{29}$       &          \\
PG 1711+336&        &                   &                   &                   &                   &                 &         \\

BH~Lyn= &3.74$^{32}$&                   &                   &                   & 13.7-13.8$^{33}$  & 75-85$^{32}$    &         \\
PG 0818+513&        &                   &                   &                   &                   &                 &         \\

BK~Lyn= &1.80$^{34}$& 0.18-0.8$^{35}$  &0.114-$>$0.185$^{36,37}$&  (0)             & 14.4-14.6$^{38}$  & 19-44$^{35}$    & 0.07-2.53\\
PG 0917+342&        &                   &                   &                   &                   &                 &         \\

AH~Men= &2.95$^{39}$& (0.82)$^{6}$      & 0.23-0.4$^{39,40}$& 0.36-0.4$^{40}$   & 13.5-13.9$^{39}$  & 0-70$^{41}$     & 0.26-5.7 \\
H0616-818&          &                   &                   &                   &                   &                 &          \\

CP~Pup= &1.47$^{42}$&0.12-0.86$^{42,43}$& 0.83-1.6$^{44,45}$&0.78-0.86$^{6,46}$ & 15.2-15.4$^{46}$  & 25-35$^{42}$    & 4.1-321 \\
Nova Pup 1942&      &                   &                   &                   &                   &                 &         \\

V348~Pup=&2.44$^{47}$& 0.35$^{48}$      &                   &                   &15.3-15.7$^{47,49}$& 80-82$^{49}$    &         \\
1H 0709-360=&        &                  &                   &                   &                   &                 &         \\
Pup1        &        &                  &                   &                   &                   &                 &         \\

\end{tabular}
\end{minipage}

$^1$ Thorstensen et al. 1991;
$^2$ Patterson et al. 1993c;
$^3$ Friedjung, Selvelli \& Cassatella 1997;
$^{4}$ Arenas et al. 2000;
$^5$ Krautter et al. 1981a;
$^6$ Warner 1995a;
$^7$ Haefner \& Metz 1985;
$^8$ Cowley et al. 1975;
$^9$ Mardirossian et al. 1980;
$^{10}$ G\"{a}nsicke et al. 1999;
$^{11}$ Wargau et al. 1982;
$^{12}$ Krautter et al. 1981b;
$^{13}$ Verbunt et al. 1997;
$^{14}$ Skillman et al. 1998;
$^{15}$ Volpi, Natali \& D'Antona 1988;
$^{16}$ Taylor et al. 1998;
$^{17}$ Huber et al. 1998;
$^{18}$ Ciardi et al. 1998;
$^{19}$ Bruch \& Engel 1994;
$^{20}$ Hellier 1993;
$^{21}$ Bonnet-Bidaud, Motch \& Mouchet 1985;
$^{22}$ DeYoung \& Schmidt 1994;
$^{23}$ Paresce et al. 1995;
$^{24}$ Balman, Krautter \& Ogelman 1998;
$^{25}$ Retter et al. 1997;
$^{26}$ Chochol et al. 1997;
$^{27}$ Retter et al. (in preparation);
$^{28}$ Shafter et al. 1990;
$^{29}$ Casares et al. 1996;
$^{30}$ Thorstensen 1986;
$^{31}$ Zwitter et al. 1994;
$^{32}$ Dhillon et al. 1992;
$^{33}$ Andronov et al. 1989;
$^{34}$ Ringwald et al. 1996;
$^{35}$ Dobrzycka \& Howell 1992;
$^{36}$ Sproats, Howell \& Mason 1996;
$^{37}$ Dhillon et al. 2000;
$^{38}$ Skillman \& Patterson 1993;
$^{39}$ Patterson 1995;
$^{40}$ Mouchet et al. 1996;
$^{41}$ Buckley et al. 1993;
$^{42}$ Duerbeck, Seitter \& Duemmler 1987;
$^{43}$ White et al. 1993;
$^{44}$ Bode, Seaquist \& Evans 1987;
$^{45}$ Williams 1982;
$^{46}$ Diaz \& Bruch 1997;
$^{47}$ Tuohy et al. 1990;
$^{48}$ Bailey 1990.
$^{49}$ Rolfe, Haswell \& Patterson 2000.

\vspace{0.25cm}
Notes to Table~1:
1. The columns (from left to right) correspond to the object (and aliases), 
orbital period in hours, white dwarf mass in solar units, distance in kpc, 
interstellar extinction in the V band, visual magnitude, inclination angle 
and mass transfer rate divided by 10$^{17}$. 
2. The original values are generally cited without any judgement. This 
is why some parameters might be clearly wrong (e.g. extreme low white dwarf 
masses of CP~Pup). A careful selection, however, has been made for the 
distance parameter. Estimates using some kind of arbitrary assumption on 
$M_{V}$ (e.g. Taylor et al. 1998) were rejected for obvious reseasons. 
3. When possible, the range of permitted magnitudes mentioned are those 
measured when permanent superhumps were detected. For the two VY~Scl
systems (TT~Ari and BH~Lyn), this means their bright states. For TV~Col, 
the mini-outburst magnitudes (Szkody \& Mateo 1984; Schwarz et al. 1988; 
Hellier \& Buckley 1993) were similarly rejected.
4. A few values of the interstellar reddening were obtained assuming 
conventional relations between $A_{V}$ and $E_{B-V}$.
5. Values in parenthesis were deduced in this work. 
6. The range of permitted inclination angles of V1974~Cyg were derived 
using the assumption that the inclination angle is similar to that of 
the ejecta plane, or to its complementary angle. It is consistent with the 
absence of eclipses in the light curves and with the presence of low 
amplitude orbital variations (Skillman et al. 1997; Retter et al. 1997). 
7. The distance estimate for V592~Cas is almost certainly underestimated.
See text for more details.
8. The upper value for the distance cited for BK Lyn is a lower limit.
Its interstellar extinction is taken to be zero to simplify the calculation.
9. The mass of AH~Men was obtained using mean values for masses of primary
white dwarfs in CVs (Warner 1995a).
10. The derived accretion rates are only rough estimates. See text for 
further details.

\end{table*}

The results are depicted in Fig.~2. The consistency of the method for 
estimating the accretion rate was discussed by Retter \& Leibowitz (1998). 
The inclination effect that is added in this study is not severe for the 
three systems discussed in that work, and the values that we obtain for 
the accretion rates of the three permanent superhump novae are rather 
similar to those deduced by Retter \& Leibowitz. We thus conclude that the 
modified version is consistent with other methods of estimating mass 
transfer rates.

\begin{figure}

\centerline{\epsfxsize=3.0in\epsfbox{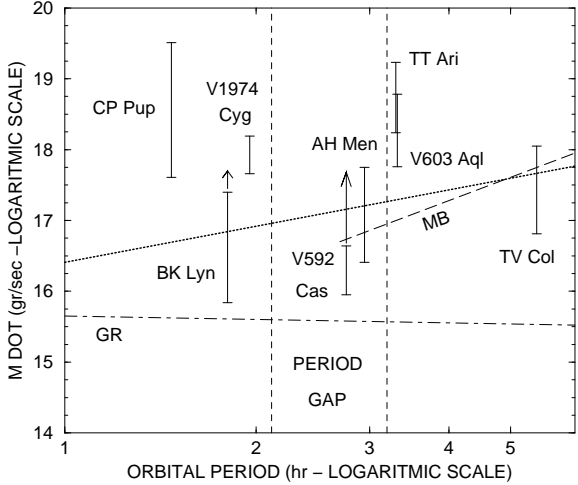}}

\caption{The location of permanent positive superhump systems in Osaki's 
(1996) diagram (Fig.~1). The two vertical dashed lines show the two 
border lines of the period gap according to Diaz \& Bruch (1997). The 
tilted dotted line separates thermally stable objects (above) and unstable 
systems (below). The bars correspond to the permitted range of the 
accretion rates derived by equation (3) with the parameters presented in 
Table~1. All permanent superhump systems have relatively high mass transfer 
rates, and are probably located above the thermal instability line, although 
some doubts exist for four systems. The arrow above the bar corresponding 
to BK~Lyn represents the fact that the distance value given by Dhillon et 
al. (2000) is a lower limit. The arrow above the permitted range of 
V592~Cas represents severe doubts on its low distance estimate. The lower 
tilted dotted-dashed line corresponds to braking by Gravitational Radiation 
(Warner 1995a). The long-dashed line that starts at P=2.7 h shows the 
calculation for Magnetic Braking (McDermott \& Taam 1989; Warner 1995a) 
using $M_{1}$=1$M_{\odot}$ and q=$M_{2}$/$M_{1}$=0.7. We note that the 
three tilted lines are model dependent. See text for more details and 
discussions.}


\end{figure}

\section{Discussion}

\subsection{Are permanent superhump systems thermally stable?}

One of our first aims in this paper was to check whether the accretion 
discs in permanent superhump systems are indeed thermally stable as was 
proposed by Osaki (1996). The results found in the previous section, and 
presented in Fig.~2 seem to support this suggestion. Among the eight 
objects in our sample with enough data, all have relatively high accretion 
rates. These values exceed in general the typical mass transfer rates in 
dwarf nova systems ($10^{15}$-$10^{16}$gr/sec), and are of the same order 
of the typical numbers in nova-likes and old novae 
($10^{17}$-$10^{18}$gr/sec -- Warner 1995a). Four systems (CP~Pup, 
V1974~Cyg, TT~Ari and V603~Aql) are located above the critical line for 
thermal instability. The permitted ranges of accretion rates of three of 
the other permanent superhumpers (BK~Lyn, AH~Men and TV~Col) straddle the 
thermal instability line and are thus consistent with their being stable. 
Note that the distance estimate for BK~Lyn given by Dhillon et al. (2000) 
is a lower limit thus implying higher values of accretion rates (see the 
corresponding arrow in Fig.~2). V592~Cas is the only permanent superhump 
system found below the instability line according to our calculations. 
Its model-dependent distance estimate (Huber et al. 1998) is very small, 
and is almost certainly incorrect as its distance to interstellar reddening
ratio is about a factor smaller than in the other objects in Table~1. The 
interstellar reddening thus implies a distance of $\sim$10 times larger, 
and therefore mass transfer rates $\sim$100 times larger (see the 
corresponding arrow in Fig.~2).

There are many observational and theoretical uncertainties in calculating 
these mass transfer rates. In fact even the location of the critical thermal 
instability line itself is controversial. Warner's (1995b) border line, 
for example, is placed about a factor of two below Osaki's (1996) critical 
line. We are looking, however, at effects of the order of $\sim$10. We thus 
conclude that the accretion discs of permanent superhump systems are most 
likely thermally stable, and that they indeed form a unique subgroup of CVs, 
with different physical parameters (namely short orbital periods and high mass 
transfer rates) from other CV subclasses. This finding further supports
the suggestion mentioned above that the distance estimate to V592~Cas quoted 
in Table~1 was underestimated, and a better measurement should raise its 
location above the critical line. Similarly, the upper limits on the 
distances of BK~Lyn, AH~Men and TV~Col are preferred to the lower values.


\subsection{Angular momentum conservation in permanent superhump systems}

Magnetic braking can account for the mean mass transfer rates of systems 
above the period gap, but is believed to occur only for systems with 
orbital periods larger than about 2.7 h. Braking by gravitational radiation 
can explain the typical accretion rates in dwarf novae below the gap, but 
there seems to be a serious problem for the short period permanent superhump 
systems (Fig.~2). The mass accretion rates estimated for these systems in 
Section~2 are $\sim$$10^{18}$gr/sec, and thus more than two 
orders-of-magnitude higher than those yielded by gravitational radiation. 
The presence of permanent superhump systems below the gap is therefore not 
understood within the current models, if their mass transfer rates 
represent mean secular values. 


One solution to this problem is to invoke an extra source of angular 
momentum loss below the period gap, the obvious candidate being magnetic 
braking. It is often stated that the secondary stars in CVs below the gap 
lack the radiative core required to anchor a magnetic field. In fact, 
single M stars (which are typical secondary stars in short orbital period 
CVs -- Smith \& Dhillon 1998), which are fully convective, do show 
magnetic activity (Fleming, Schmitt \& Giampapa 1995). Whilst invoking 
such braking would preclude certain explanations of the period gap itself, 
which rely on the cessation of magnetic activity (e.g. Verbunt 1984), it 
would allow a natural explanation for the presence of high accretion rates 
in systems with orbital periods below the period gap. Therefore, we 
suggest that the mere existence of permanent superhump systems below the 
gap might be an argument that magnetic braking does not cease at the gap. 

A different solution to the problem is provided by the possibility of
mass transfer cycles. If the permanent superhump phase is very short lived, 
and systems below the gap spend most of their time as SU~UMa systems, 
perhaps even hibernating for a while, then the mean mass transfer rate 
might not exceed the Gravitational Radiation values. An argument against 
such a possibility is that King et al. (1996) could not produce self 
sustaining mass transfer cycles in short period systems. However, as we 
shall show in the next section, there may be emerging observational 
evidence that nova explosions drive such cycles below the gap.

\subsection{Nova cycle scenarios for non-magnetic short orbital period CVs}

Post-novae with orbital periods above the gap usually return to their 
pre-outburst brightness within a few decades (Section~1.4). So far only 
two non-magnetic novae below the period gap, CP~Pup and V1974~Cyg, have 
been found. Both have permanent superhumps in their light curves suggesting 
that their discs are thermally stable. Table~2 presents the pre-outburst 
and post-eruption magnitudes of the two systems. The observations are 
consistent with the two post-novae being brighter than their progenitors. 
Moreover, the pre-nova magnitude of V1974~Cyg implies that its disc was 
thermally unstable and therefore the progenitor of the nova should have 
been an SU~UMa system (Retter \& Leibowitz 1998). Thus, it is the first 
example of a nova that has changed the thermal stability state of its 
accretion disc\footnote{It is unknown whether V446~Her and GK~Per, the 
two novae that show dwarf nova outbursts in their light curves 
(Section~1.4), had nova-like stages before or after their nova outbursts. 
Note that it is unclear whether accretion discs survive nova eruptions.}. 
This transition might be explained by a temporary change in the mass transfer
rate due to the hot post-nova white dwarf irradiating the secondary star
or the disc itself. Alternatively, V1974~Cyg is the first direct evidence 
for mass transfer cycles, in this case driven by the nova explosion. 
CP~Pup is a candidate for the same behaviour. The nova cycle scenarios 
proposed for long orbital period CVs (Section~1.4) could thus be applicable 
to the short period systems, but possibly with some minor modifications.

\begin{table}
\caption{Pre-eruption and post-outburst magnitudes of the two permanent 
superhump novae below the period gap}
\begin{tabular}{@{}lcccc@{}}


Object    & year of & pre-nova      & post-nova      & difference \\
name      & outburst& magnitude     & magnitude      &            \\


CP~Pup    & 1942    & $>17^{1}$     & 15.3$^{2}$     & $>1.7$     \\
V1974~Cyg & 1992    & 21$^{3,4}$    & 16.5--?$^{5}$  & 4.5--?     \\


\end{tabular}

$^1$ Warner 1995a;
$^2$ Patterson \& Warner 1998;
$^3$ Pavelin et al. 1993;
$^4$ Retter \& Leibowitz 1998;
$^5$ Goranskiy 2000,







\end{table}

\section{Summary and conclusions}

Our results can be summarized as follows:\\
1. We established the idea that the accretion discs in permanent superhump 
systems are thermally stable.\\
2. If the high values of accretion rates found in the short orbital 
period permanent superhumpers represent their secular mean values, then 
there is a problem with the mechanism that removes angular momentum from 
the systems. It might be solved by extending the magnetic braking mechanism 
to below the gap, or by invoking another way to lose angular momentum in 
these CVs. Alternatively, systems below the period gap might have mass 
transfer cycles, and the permanent superhump stage should be short compared 
with the full CV cycle.\\
3. We suggest that nova cycle scenarios similar to those proposed for the 
long orbital period CVs should be applied to the short period systems. 
However, they would be slightly modified if the superhumps observed in the 
post-novae below the gap represent a true long-term change in the mass 
transfer rate rather than a transient increase in the disc luminosity due 
to irradiation by the hot white dwarf. The observations of the two 
non-magnetic novae below the gap seem to support mass transfer cycles 
driven by the nova outburst.

\section{Acknowledgments}

Hans Ritter, the referee, is acknowledged for many valuable comments.
We also thank Dina Prialnik, Elia Leibowitz, Mike Shara, Joe Patterson, 
Rob Jeffries, Michael Friedjung, Coel Hellier and Sandi Catalan for 
fruitful discussions. TN is a PPARC advanced fellow. AR is supported by 
PPARC.

\end{document}